\documentclass[12pt]{article}
\usepackage{cite,graphicx,amsmath,amssymb,bbm}
\usepackage{epsf}

\newcommand{\be}{\begin{equation}}  
\newcommand{\ee}{\end{equation}}  
\newcommand{\bea}{\begin{eqnarray}}  
\newcommand{\eea}{\end{eqnarray}}

\newcommand{\ol}[1]{\overline{#1}}

\begin{document}

\thispagestyle{empty}
\vspace*{.2cm}
\noindent
HD-THEP-06-12 \hfill 18 July 2006
\\
\noindent
OUTP-DR-06 01P

\vspace*{1.0cm}

\begin{center}
{\Large\bf The Ubiquitous Throat}
\\[1.5cm]
{\large A.~Hebecker$\,^a$, and J.~March-Russell$\,^b$}\\[.5cm]
{\it ${}^a$ Institut f\"ur Theoretische Physik, Universit\"at Heidelberg,
Philosophenweg 16 und 19\\ D-69120 Heidelberg, Germany}
\\[.3cm]
{\it ${}^b$ Rudolf Peierls Centre for Theoretical Physics, University 
of Oxford, 1 Keble Road\\ Oxford OX1 3NP, UK}
\\[.4cm]
{\small\tt (a.hebecker@thphys.uni-heidelberg.de} {\small and}
{\small\tt j.march-russell1@physics.ox.ac.uk\,)}
\\[1.0cm]

{\bf Abstract}
\end{center} 
We attempt to quantify the widely-held belief that large hierarchies 
induced by strongly-warped geometries are common in the string theory
landscape. To this end, we focus on the arguably best-understood subset 
of vacua -- type IIB Calabi-Yau orientifolds with non-perturbative K\"ahler 
stabilization and a SUSY-breaking uplift (the KKLT setup). Within this 
framework, vacua with a realistically small cosmological constant are 
expected to come from Calabi-Yaus with a large number of 3-cycles. For 
appropriate choices of flux numbers, many of these 3-cycles can, in general, 
shrink to produce near-conifold geometries. Thus, a simple statistical 
analysis in the spirit of Denef and Douglas allows us to estimate the 
expected number and length of Klebanov-Strassler throats in the given 
set of vacua. We find that throats capable of explaining the electroweak 
hierarchy are expected to be present in a large fraction of the landscape 
vacua while shorter throats are essentially unavoidable in a statistical 
sense.

\newpage
\section{Introduction}
Since the seminal work of Giddings, Kachru and Polchinski~\cite{
Giddings:2001yu} following on from the foundational papers 
of Refs.\cite{earlywarp} and \cite{Verlinde:1999fy},
it has become common knowledge that strongly warped 
regions or throats are a natural feature of type IIB flux compactifications
(see~\cite{Grana:2005jc} for a recent review). Moreover, thanks to the KKLT 
construction~\cite{Kachru:2003aw}, the very same class of models has become 
the nucleus of the large and growing collection of metastable de-Sitter 
vacua of string theory (known with a varying degree of rigour) which are 
generally referred to as the `string theory landscape' \cite{landscape}. Following the 
line of thought developed by Douglas and collaborators~\cite{Douglas:2003um,
Denef:2004ze,Denef:2004cf}, it is then natural to attempt to link the 
presence of throats quantitatively to the assumption that we live in one of 
the numerous type IIB orientifold models with 3-form flux. It is the aim of 
the present paper to understand to which extent throat and multi-throat 
geometries can be considered a prediction of the type IIB landscape proposal. 

To be specific, we will focus on the oldest and arguably simplest 
situation~\cite{Kachru:2003aw} in which, given a model where all 
complex structure moduli are stabilized by 3-form flux, the single K\"ahler 
modulus is stabilized non-perturbatively by gaugino condensation or D3-brane 
instantons. We have every reason to expect that our conclusions, which will 
mainly be related to the distribution of 3-form flux quanta on the various 
3-cycles, remain valid if K\"ahler moduli are stabilized by the interplay of 
perturbative and nonperturbative physics~\cite{Balasubramanian:2004uy, 
Balasubramanian:2005zx} or even in an entirely perturbative 
fashion~\cite{Saltman:2004sn,vonGersdorff:2005bf}. Similarly, we do not 
expect our conclusions to be affected by the modifications and extensions 
of the stabilization mechanism required in situations with more than one 
K\"ahler modulus (see e.g.~\cite{Balasubramanian:2005zx}). 

Given that all geometric moduli are stabilized in a supersymmetric AdS 
vacuum as described above, we assume, following KKLT~\cite{Kachru:2003aw}, 
that a small supersymmetry breaking effect, such as the presence of 
anti-D3-branes in one of the warped regions, uplifts this vacuum to a 
de-Sitter vacuum with realistic cosmological constant. We choose to focus on 
this (by now classic) scenario since the metastability of such uplifted vacua 
is essentially guaranteed in the limit of a parametrically small AdS 
cosmological constant before the uplift. We will comment on this in more 
detail below. However, we emphasize again that our decision to be so 
restrictive in our choice of models is motivated solely be the desire to 
keep the non-essential parts of our analysis short and simple. We expect 
that the distribution of throats emerging from our analysis will be similar 
in a much wider class of flux vacua. 

Given the above considerations, we focus on the distribution of throats in 
type IIB Calabi-Yau orientifolds with a large number of 3-cycles and under
the restriction that the total flux superpotential $W_0$ at the SUSY minimum 
is parametrically small. It is natural to expect that, as a result of the 
random choice of a large number of independent flux quanta for the various 
3-cycles, some of these 3-cycles will automatically carry only a small
number of flux. If this occurs for a 3-cycle that can shrink to produce 
a conifold singularity~\cite{Candelas:1989js} (which may be the generic 
situation) and if the flux carried by the dual cycle is not small, a 
Klebanov-Strassler throat~\cite{Klebanov:1998hh,Klebanov:2000hb} with 
an exponentially large hierarchy develops. This is the naive expectation and 
at the same time our main result: The more detailed analysis described 
in bulk of the paper confirms that one has to expect large hierarchies of 
scales~\cite{Randall:1999ee,Verlinde:1999fy,Goldberger:1999uk} and multiple 
throats~\cite{smallnumbers,Barnaby:2004gg,Cascales:2003wn} in generic 
orientifold models of the landscape. 

At a more technical level, we will replace the above heuristic argument 
about `accidentally' small 3-cycles by the quantitatively well-known fact
that vacua accumulate near conifold points~\cite{Denef:2004ze,
Giryavets:2004zr,Conlon:2004ds,Eguchi:2005eh}. If many such conifold points 
are present in the moduli space of a given Calabi-Yau, the probability of 
being far away from any of them becomes extremely small. In this sense, the 
presence of throats becomes a prediction of the given branch of the string 
theory landscape.

\section{The relevant set of vacua}
Following~\cite{Denef:2004ze}, we consider the orientifold limit of an 
F-theory compactification based on a four-fold with Euler number $\chi_4$. 
The 3-form flux on this orientifold can be quantified by a flux vector
$N\in {\mathbb Z}^{2K}$. Its dimension is given by $2K=4(h^{2,1}_-+1)$, where 
$h^{2,1}_-$ is the number of complex structure moduli.\footnote{
The 
index `$-$' is used since, as an alternative to the F theory construction, 
one may think of orientifolding a smooth Calabi-Yau 3-fold to obtain a
given model. The relevant cycles are those which are odd under the 
orientifold projection. Note that our $K$ is $K/2$ in the notation 
of~\cite{Denef:2004ze}
}
Allowing for a contribution $N_{D3}$ to the total D3-brane charge from 
freely moving D3 branes $(N_{D3}>0)$ or anti-D3-branes $(N_{D3}=-N_{\ol{D3}} 
<0)$, the tadpole cancellation condition reads
\be
\frac{\chi_4}{24}=\frac{1}{2}N^T\Sigma N+N_{D3}\,,
\qquad\mbox{where}\qquad
\Sigma\equiv\left(\begin{array}{cc}0&\mathbbm{1}\\ \mathbbm{1}&
0\end{array}\right)\,.
\ee
Assuming that the orientifold planes of the model preserve the same 
supersymmetry as D3 (rather than anti-D3) branes and focusing on 
supersymmetric vacua,\footnote{
To 
be more precise, these are SUSY-breaking no scale vacua which turn into 
supersymmetric AdS vacua once non-perturbative K\"ahler stabilization is 
taken into account.
}
one requires $N_{D3}>0$. The flux vector is then subject to the constraint 
\be 
\frac{1}{2}N^T\Sigma N\equiv L\le L_* \equiv \frac{\chi_4}{24}\,.
\ee
The number of SUSY vacua available in this situation was estimated in~\cite{ 
Denef:2004ze} to be 
\be
{\cal N}_{susy}(L\le L_*) \sim \frac{L_*^K}{K!}\,.\label{nsusy}
\ee

We are, however, interested specifically in realistic vacua (i.e. vacua with 
small positive cosmological constant) originating from non-perturbative 
K\"ahler stabilization combined with an anti-D3-brane uplift based on a 
small positive $N_{\ol{D3}}$. To ensure stability, no freely moving D3 branes 
should be present in this construction. Furthermore, to guarantee perturbative 
control and a sufficiently long lifetime of the metastable anti-D3-brane 
configuration at the bottom of the throat, we require $N_{\ol{D3}}\leq 
N_{\ol{D3}\,,\,max}$. An upper bound on $N_{\ol{D3}\,,\,max}$ is provided by 
the classically-allowed decay process studied in~\cite{Kachru:2002gs} which 
limits the range of metastability to $N_{\ol{D3}}< 0.08 M$ where $M$ is the 
RR-flux quantum. As in our counting of flux vacua we scan over flux quanta up 
to $L_*$ we take the parametric dependence $N_{\ol{D3}\,,\,max}\ll L_*$. 

The number of such `uplifted' vacua can be estimated by an appropriate 
modification of Eq.~(\ref{nsusy}):
\be 
{\cal N}_{uplift}={\cal N}_{susy}(L_*<L\le L_*+N_{\ol{D3}\,,\,max}) 
\sim N_{\ol{D3}\,,\,max}\,\frac{L_*^{K-1}}{(K\!-\!1)!}\
\sim \frac{L_*^K}{K!}\,,
\ee
where we have Taylor expanded in $N_{\ol{D3}\,,\,max}$ and dropped irrelevant 
non-exponential factors in the last expression to simplify the final formula. 
Thus, ${\cal N}_{uplift}$ has the same parametric behaviour as ${\cal 
N}_{susy}$. Clearly, the cosmological constants of these uplifted vacua can 
have both signs and vary widely in their value. The source for this 
variation is the flux superpotential 
\be
W=\int G_3\wedge\Omega\,,
\ee
which provides a negative contribution $\sim |W_0|^2$ in each vacuum, to be 
(under- or over-) compensated by the uplift $\sim N_{\ol{D3}}$. Given that 
both Re$\,W_0$ and Im$\,W_0$ depend linearly on the flux vector, one expects 
a uniform distribution of vacua in the central region of the complex $W_0$ 
plane. This, in turn, implies a uniform distribution of $|W_0|^2$ on the 
positive real axis. If we ignore any moderate volume suppression and 
non-exponential factors depending on $L_*$ and $K$, the maximal size of 
$W_0$ is string scale (i.e. ${\cal O}(1)$ in our units). Thus, the 
probability that the negative contribution $\sim |W_0|^2$ compensates a 
fixed positive $V_{uplift}$ with enough precision to come close to the 
observed cosmological constant $\Lambda$ is approximately equal to $\epsilon 
\sim\Lambda 
\sim 10^{-120}$. ($V_{uplift}$ should be small enough to allow perturbative 
control but large enough to avoid any peculiarity that the $W_0$ 
distribution might have very close to the origin.) We conclude that one 
needs geometries with ${\cal N}_{uplift} \sim 10^{120}$ to have ${\cal O}(1)$ 
probability for a (cosmologically) realistic vacuum to exist. 

We are interested in an estimate for the lowest $K$ that is consistent within 
the present framework. Thus, we choose $L_*=\chi/24$ as large as possible 
(within the presently known set of Calabi-Yau 4-folds) and estimate $K$ on 
the basis of 
\be
\frac{L_*^K}{K!}\sim \frac{1}{\epsilon}\,.\label{keq}
\ee
Ignoring non-exponential factors in Stirling's formula and assuming 
that $\log(eL_*/K)\simeq {\cal O}(1)$, one finds $K_0\sim \log(1/\epsilon)$.
A better estimate of $K$ follows from replacing $K!$ with $(K_0/e)^K$ 
on the lhs of Eq.~(\ref{keq}) leading to
\be
K\sim \frac{\log(1/\epsilon)}{\log[eL_*/\log(1/\epsilon)]}\,.
\ee
For $L_*\sim 10^4$ (see, e.g.,~\cite{Klemm:1996ts}), we find $K\sim 60$,
corresponding to $h^{2,1}_-\sim 30$. 

It is important to keep in mind that this is just a lower bound and 
that, most probably, the number of cycles of a `typical' flux 
compactification with realistic cosmological constant is significantly 
larger. For example, one might say that with typical Calabi-Yaus having 
$h^{2,1}\sim 100...200$ (see e.g.~\cite{Candelas:1994hw,Avram:1996pj, 
Denef:2004dm})\footnote{
Extreme 
cases of $h^{2,1}\sim 500$ are known, see, eg, 
{\tt http://hep.itp.tuwien.ac.at/\~{}kreuzer/CY/}
}  
we can very naively expect that $h^{2,1}_{-}\sim 50...100$ is typical.
Moreover, the scaling of the number of vacua (as implied by Eq.~(\ref{keq}) 
for $K<L_*$) suggests that CY's with the largest
possible value of $K$ are exponentially preferred in that they allow a far 
greater number of flux vacua. For example, ${\cal N}_{susy}(K=200)/{\cal 
N}_{susy}(K=60) \sim 10^{270}$ for fixed $L_*\sim 10^4$.
 
Given our ignorance of the model describing our vacuum as well as of the 
mechanism choosing physical compactification manifolds, we keep 
$h^{2,1}_{-}$ and $\chi_4$ (or equivalently $K$ and $L_*$) as unknown 
parameters with the order of magnitude given above.

The complex structure moduli spaces of such complicated orientifold models 
have not been analyzed in detail. It is clear that they will contain various
regions where certain 3-cycles blow up or shrink to zero size. We will
henceforth ignore the former `large complex structure' regions although
they might, in fact, be interesting and important to study. Instead, in 
this paper we focus on the singularities arising when one or more of the 
3-cycles shrink. We want to argue that, in many cases, these singularities 
are `nodes' or `ordinary double points', which are particularly common 
singularities of complex varieties. Nodal 3-folds arise naturally in 
algebraic topology, one of the prominent examples being the various 
singular limits of the quintic hypersurface in 4d complex projective space. 
In this specific case, it is known that the `generic' singular space has a
single node~\cite{Lefschetz}. From the perspective of the Calabi-Yau 3-fold 
defined in this way, such a point corresponds to a conifold 
singularity~\cite{Candelas:1989js} (which develops as one of the 3-cycles 
shrinks). Furthermore, a large set of smooth Calabi-Yaus is linked by 
conifold transitions into a `Web' (including, in particular, the 
quintic)~\cite{Candelas:1989ug}. In each case, the singular intermediate 
situation is approached from one side of the transition by the shrinking of 
a number of 3-cycles with $S^3$-topology (the conifold limit mentioned above)
\cite{Curio:2000sc}.  From this we conclude that the conifold limit is a common
(possibly the generic) way in which a 3-cycle of a Calabi-Yau shrinks. More
specifically, we assume in the following that an ${\cal O}(1)$ fraction of the possible 
limits of shrinking 3-cycles of the models under consideration correspond to 
conifold points. It is an interesting question (which goes beyond the scope 
of this work) to understand for how many of the known Calabi-Yau orientifolds 
this assumption holds. 

The distribution of flux vacua in the vicinity of such conifold 
points has been analyzed at least for certain simple examples. It has been 
found that vacua accumulate near these points. This can be understood 
intuitively by recalling that the distance $|z|$ from a conifold point is 
given by~\cite{Giddings:2001yu}
\be 
z\sim \exp(-2\pi P/g_sM)\,.
\ee
Here $z$ is the complex structure modulus corresponding to the shrinking 
3-cycle while $M$ and $P$ are the numbers of flux quanta on the conifold 
cycle and on its dual. It is then clear that a smooth distribution of flux 
quanta can lead to a strong enhancement of the number of vacua with 
exponentially small $z$. 

More specifically, it was shown in~\cite{Denef:2004ze} that, in a given 
model with one conifold point at $z=0$ and a fixed tadpole constraint 
$L_*$, the fraction of vacua with conifold cycle smaller than $|z|$ decays 
as 
\be 
{\cal N}(z)\sim \frac{1}{\log(1/|z|)}\label{enh}
\ee
for $z\to 0$. Clearly, this implies an enhancement of the number of vacua 
very close to the conifold point relative to naive expectations that one 
might have on the basis of the canonical measure on ${\mathbb C}$.

\section{Stability issues}
Before turning to our main interest, the distribution of throats, we would 
like to address the stability of the above set of vacua after uplifting. 

Naively, one might expect the following situation: We focus on the complex 
structure moduli $z_i$ and the dilaton modulus $\tau$. A generic flux 
induced modulus mass is ${\cal O}$(1) in string units (if we ignore any 
volume suppression $\sim{\cal O}(\mbox{few})$). Making the vacuum value 
$W_0$ of the superpotential $W$ parametrically small by tuning fluxes, we 
obtain a parametrically small cosmological constant $\Lambda_{\rm AdS}\sim 
-|W_0|^2$. Consider now the scalar potential
\be
V=e^K(K^{a\ol{b}}D_aW\ol{D_bW}-3|W|^2)\,\label{sgp}
\ee
near the supersymmetric point, where $D_aW=0$ and $W=W_0$ (the index $a$ 
labels the moduli $\phi_a=(\tau,z_i)$). The scalar mass matrix near this 
point gets an ${\cal O}(1)$ contribution from the first term, which is 
positive definite since the inverse K\"ahler metric has this feature. It 
also gets (potentially negative) contributions 
\be
\sim -e^KK_{a\ol{b}}|W_0|^2\qquad\mbox{and}\qquad\sim -e^K(D_aD_bW)
\bar{W}_0
\ee
from the second term (where we again used the fact that $D_aW$ vanishes in 
the vacuum). Thus, in the generic case, all masses should be positive and 
${\cal O}(1)$ if $W_0$ is parametrically small. (This can also be argued
by appealing to the known stability of supersymmetric vacua in combination 
with the Breitenlohner-Freedman bound~\cite{Breitenlohner:1982bm}: If all 
mass squares are ${\cal O}(1)$ and $\Lambda_{\rm AdS}$ is small, all mass 
squares must be positive.)

However, in the conifold limit of the one-modulus case analyzed 
in~\cite{Denef:2004ze}, this naive expectation was found to be 
violated and tachyonic directions (implying the danger of physical 
instabilities after uplifting) were found to be generically present. 
At the same time, it was argued that this problem will not persist in 
models with more than one complex structure modulus. We agree with this
expectation and we would like to supply an explicit argument in its favour:

Ignoring the ${\cal O}(1)$ prefactor $e^K$ and using the parametric smallness 
of $W_0$, the second-order expression for the supergravity scalar potential, 
Eq.~(\ref{sgp}), takes the form 
\be
V\sim \delta\phi_a\,W_{ab}\,K^{b\ol{c}}\,\ol{W}_{\ol{c}\ol{d}}\,\delta 
\ol{\phi}_{\ol{d}}\label{mm}
\ee
near the vacuum. Here $\delta\phi_a$ is the deviation of $\tau$ (for $a=1$) 
or any of the complex structure moduli (for $a>1$) from its vacuum value. 
This is, of course, just the familiar rigid-SUSY expression. It is obvious 
from Eq.~(\ref{mm}) that a parametrically small eigenvalue of $W_{ab}$ leads 
to a light scalar field which, taking into account the full supergravity 
expression (including non-vanishing $W_0$) and the uplift, entails the 
risk of a tachyonic direction. Indeed, such a small eigenvalue arises in 
the one-complex-structure-modulus case near the conifold point: The explicit 
form of the superpotential
\be
W=A(z)+\tau B(z)\,,
\ee
implies $W_{11}=0$ and the singular behaviour of the integral over the 
dual conifold cycle, 
\be
\int_B\Omega=\frac{z}{2\pi i}\log(z)\,+\,\mbox{holomorphic}\,,\label{lnz}
\ee
implies $W_{22}\sim 1/z$. The 2$\times$2 matrix $W_{ab}$ then develops a 
parametrically small eigenvalue by the usual see-saw mechanism, which 
makes the tachyonic direction found in~\cite{Denef:2004ze} possible. 

The situation changes drastically in the case of two or more complex 
structure moduli. Let $W_{ab}$ be an $n\times n$ matrix and let $a=n$ 
correspond to the conifold modulus $z$ of Eq.~(\ref{lnz}). While it is 
still true that $W_{11}=0$ and $W_{nn}$ is parametrically large, this no 
longer implies the existence of a small eigenvalue. This can be seen by 
considering the characteristic equation
\be
\mbox{det}(W_{ab}-\lambda\delta_{ab})=0\,.
\ee
Clearly, the largeness of $W_{nn}$ implies the existence of a large 
eigenvalue $\lambda\simeq W_{nn}$. Any further eigenvalue, however, has to 
solve the equation 
\be
\mbox{det}\left(W_{ab}-\lambda\delta_{ab}\big|_{\{a,b=1\ldots n-1\}}\right)
=0\,,
\ee
approximately. The solutions are simply the eigenvalues of an $(n-1)\times 
(n-1)$ matrix with vanishing upper-left element, which is otherwise generic. 
For $n>1$, neither of these eigenvalues is generically small. Thus, we have 
no reason to expect that the problem of a tachyonic direction observed in the 
case of a single complex structure modulus will persist. 

Independently of the above, it may also be useful to observe that the 
special feature $W_{11}=0$ of the KKLT construction is not generic and can
easily be avoided, e.g., by including gaugino condensation on stacks of D3 
branes at singularities.\footnote{
Related 
discussions of the stability of the KKLT construction can be found, e.g., 
in~\cite{Brustein:2004xn,Choi:2004sx}.
} 

Our main conclusion for the following is that the difficulties 
observed in the one-modulus case do not represent an argument against the 
existence of many uplifted near-conifold vacua of fluxed multi-modulus 
Calabi-Yaus.

\section{The distribution of throats}
We now turn to our main point, which is the interplay between the expected 
large number of 3-cycles (and hence of potential conifold singularities) 
and the enhancement of the number of vacua in the vicinity of each of those 
singularities.

Consider the complex structure moduli space of an orientifold model with 
$\sim K$ $\mbox{3-cycles}$, as discussed in the previous section. We assume, 
motivated by the example of the quintic and the `Web of Calabi-Yaus', that
an ${\cal O}(1)$ fraction of these cycles produce, when they shrink, conifold 
singularities. Furthermore, we excise all large complex structure regions, 
ending up with a compact moduli space of complex dimension $K/2$. The various 
conifold points are described by ${\cal O}(K)$ subspaces of complex 
co-dimension one which, in general, intersect each other.

Let us first focus on one of these conifold points (more properly: on one 
of the subspaces along which a certain conical singularity persists) and 
parameterize the moduli space such that the coordinate $z_i$ characterizes 
the shrinking cycle. Making use of the distribution of vacua near a conifold 
singularity implied by Eq.~(\ref{enh}), we expect that a randomly chosen 
flux vacuum will have probability 
\be
p_i(|z_i|)\simeq \frac{1}{c_i\log(1/|z_i|)} \label{prob}
\ee
to be less than $|z_i|$ away from the conifold point under consideration.

The real constant $c_i$ is related to the detailed distribution of vacua 
away from the conifold point and to the ambiguities which arise in excluding 
the large complex structure regions. To see this, assume for simplicity 
that we have excised the region $|z_i|>|z_{i\,,\,max}|$. Clearly, away 
from the small-$z_i$ region $p_i$ has some more complicated functional form 
(not explicitly known in general) and it has to satisfy the normalization 
condition 
\be
p_i(|z_i|)\to 1\qquad\mbox{for}\qquad |z_i|\to|z_{i\,,\,max}|\,.
\ee
All that we can infer from Eq.~(\ref{enh}) is that $p_i$ is {\it 
proportional} to $1/\log(1/|z_i|)$ at small $z_i$; the normalization is 
inextricably linked to the behaviour of $p_i$ at $|z_i|\sim|z_{i\,,\,max}|
\sim{\cal O}(1)$. What is worse, $c_i$ is in general a function of the other 
complex structure moduli, $c_i=c_i(z_1,\dots,z_{i-1},z_{i+1},\ldots, 
z_{K/2})$ (which we ignored in the above), thereby making a detailed analysis 
of the full probability distribution highly non-trivial. Thus, all that we 
can do at the moment is to assume that the various $c_i$ do not vary too 
rapidly and are not parametrically large or small (for which there is no 
obvious reason). We will parameterize our ignorance assigning a universal 
unknown value of the order of one to all these coefficients, $c_i=c$. 

Let us now recall that, if $z_i$ is stabilized near zero, a strongly warped 
region or throat with a hierarchy of mass scales 
\be
h_i \sim |z_i|^{-1/3}
\ee
between the Klebanov-Strassler region (IR end) and the Calabi-Yau region 
(UV end) develops~\cite{Giddings:2001yu} (see also~\cite{DeWolfe:2002nn}). 
We conclude from the above that the probability for finding a throat with a 
hierarchy larger than $h_i$ is
\be
p_i(h_i)\simeq\frac{1}{3c_i\log\,h_i}\,.
\ee
Intuitively, this characterizes the probability for being, in the given 
moduli space, within a slice of a certain thickness that surrounds the 
hypersurface defined by $z_i=0$. 

Assuming that these probabilities are uncorrelated for the various 
conifold hypersurfaces (i.e. for the various $z_i,\, i=1\ldots K$), we can 
estimate the probability for finding precisely $n$ throats with hierarchy 
larger than $h_*$. It is given by the probability for being inside $n$ of the 
$K$ slices and outside the remaining $K-n$ slices, multiplied by a 
combinatorial factor for choosing inside which slices to be:
\be
p(n,h>h_*|K)\sim \binom{K}{n}\,p^n(1-p)^{K-n}\qquad\mbox{with}\qquad
p\equiv\frac{1}{3c\log h_*}\,.
\ee
The fact that this `multi-throat probability' is given simply by a 
binomial distribution with parameters $K$ and $p$ represents one of our main 
results (or, given the various assumptions above, our main conjecture). 
Many interesting and potentially phenomenologically relevant questions 
can now be addressed.

For example, given a certain hierarchy factor $h_*$, we can inquire about the 
expected number of throats with a larger hierarchy. It is given by the 
well-known mean of the binomial distribution,
\be
\ol{n}(h>h_*|K)=\sum_{n=0}^K n\,\binom{K}{n}\,p^n(1-p)^{K-n}=Kp=\frac{K}
{3c\log h_*}\,.\label{nbar}
\ee
The crucial but certainly not unexpected point here is that $\bar{n}$ goes
to zero very slowly as $h_*$ grows. The variance of $n$, again a familiar 
result, is
\be
\mbox{var}(n)=\sum_{n=0}^K(n-\ol{n})^2\,\binom{K}{n}\,p^n(1-p)^{K-n}=
Kp(1-p)\,,
\ee
which is very close to $\ol{n}$ for $p\ll 1$. Thus, the expected number of 
throats is
\be
\ol{n}\pm\sqrt{\ol{n}}\,,
\ee
with $\ol{n}$ as given in Eq.~(\ref{nbar}).

For a given $K$ the probability that at least one throat has hierarchy 
exceeding some specified $h_*$ is given by
\be
P(h>h_*|K) = \left(1-\frac{1}{3c\log h_*}\right)^K \left[ \left(1+\frac{1}
{3 c \log h_* -1}\right)^K -1\right]\,.\label{atleast}
\ee
Figure~1 shows this function against $\log h_*$ for $c=1$ and $K$
taking the values $60$ and $200$. It is noteworthy that there is a very 
slow decrease of the probability with throat length,
and that at 50\% likelihood there exist throats of hierarchy greater than
$\exp(28)\sim 10^{12}$, and $\exp(95)\sim 10^{41}$ as
$K$ varies from 60 to 200. Not surprisingly, Eq.~(\ref{atleast}) coincides 
with Eq.~(\ref{nbar}) if $K/3c\log(h_*)\ll 1$.
  
\begin{figure}
\begin{center}
\includegraphics[width=10cm]{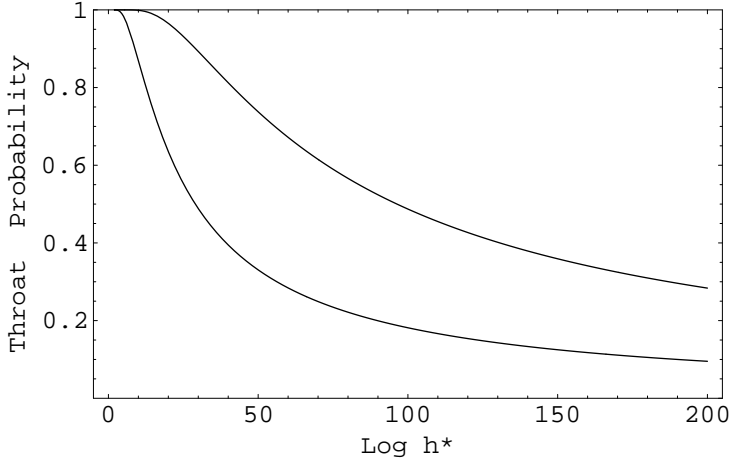}
\caption{Probability that at least one throat has hierarchy $h>h_*$ as a 
function of $\log h_*$. Reading from bottom to top the curves correspond to 
the choices $K=60$ and 200 (both taking $c=1$).}
\label{pot}
\end{center}
\end{figure}

Another interesting quantity is the hierarchy $h_1$ of the longest expected
throat. A simply estimate of this quantity is provided by solving 
\be
\ol{n}(h>h_1|K)\sim 1
\ee
for $h_1$. The result is
\be
\log h_1\sim\frac{K}{3c}\,.\label{h1}
\ee
Alternatively, we can ask for which $h_1$ the one-throat-probability is 
maximized,
\be
\frac{d}{d\,h_1}\,p(1,h>h_1|K)=0\,.
\ee
The result is consistent with Eq.~(\ref{h1}). Yet another way to state 
the same problem is to ask, at fixed $h_1$, for the value of $K$ which gives 
the maximal value for $p(1,h>h_1|K)$. Again, the resulting relation of $h_1$ 
and $K$ is approximately that of Eq.~(\ref{h1}). 

Furthermore, a very simple but important quantity is the probability of having 
no throat with a hierarchy larger than $h_*$,
\be
p(0,h>h_*|K)\simeq (1-p)^K \simeq \exp\left(-\frac{K}{3c\log h_*}\right)
\,.\label{p0}
\ee
As expected, this is a very small number for large $K$ and not too large
hierarchies. We consider this together with the expected number of throats, 
Eq.~(\ref{nbar}), the expected hierarchy of the longest throat, 
Eq.~(\ref{h1}), and the one-throat-probability of Fig.~\ref{pot} to be the 
main results of this section. 

Finally, using Eq.~(\ref{atleast}) it might also be possible to gain
information on $K$ independent of the fine-tuning of the cosmological constant
by conditioning on the existence of an electro-weak throat with IR 
scale $\sim$TeV. Using Bayes' theorem the conditional probability 
distribution for $K$ given that there exists at least one throat with $h\geq 
h_{EW}$ is
\be
P(K|n(h_{EW})\geq 1) = \frac{P(h>h_{EW}|K) P(K)}{\Sigma_{K'=1}^{K_{max}} 
P(h>h_{EW}|K') P(K')} .\label{ewprior}
\ee
If we conservatively assume a flat prior distribution for $K$, $P(K) = 
1/K_{max}$ and take as an illustrative example $K_{max}=200$ and $c=1$, then 
a numerical evaluation of Eq.~(\ref{ewprior}) leads to an {\it a posteriori} 
mean ${\bar K} \sim 124$.

\section{Possible phenomenological implications}
To discuss possible phenomenological implications, we have to quantify the 
expected hierarchies $h$, which depend crucially on the number of cycles $K$. 
Since, at the fundamental level, we are ignorant about $K$ and, moreover, $K$
appears in the combination $K/3c$ (with an unknown ${\cal O}(1)$-constant 
$c$), we take the following pragmatic approach:

We consider two scenarios, one conservative and one more favourable: In the 
conservative case, we choose $K=60$ (roughly the minimal value consistent 
with fine-tuning $\Lambda$) and 
$c=3$, such that the relevant combination of these two parameters takes the 
low value $K/3c\simeq 7$. In the favourable case, we choose $K=200$ 
(consistent with typical Calabi-Yau values, maybe somewhat at the high side, 
but not extreme). Together with $c=1/3$ this gives the high value 
$K/3c\simeq 200$. 

In the conservative case, Eq.~(\ref{h1}) implies that the longest throat 
typically has a hierarchy $\sim 10^3$. This clearly also means that,
specifying a minimal hierarchy $10^3$, we expect about one throat with 
a hierarchy above that value. We can also infer that we have to expect about 
3 throats with hierarchy 10 or larger. Even though these numbers are not 
very impressive, they clearly imply that dynamically generated scales 
of $\sim 10^{-3}M_P$ are natural in the present branch of the landscape. 
The above short throats can play an important role in inflation or 
simply to ensure a small (and hence perturbatively controlled) anti-D3-brane 
uplift. Thus, even though no spectacular low-energy effects can be predicted 
in this conservative setting, moderate throats are indeed `ubiquitous'. 

What is maybe more impressive is the small statistical price that one pays 
for having a moderately long throat. For example, the expectation value for 
the number of throats with hierarchy above $10^6$ is $0.5$. In other words, 
low scale SUSY in the KKLT setting (see e.g.~\cite{Choi:2005ge}) is perfectly 
plausible and does not require any extra fine-tuning. Even more, demanding a 
hierarchy of $10^{13}$ or higher, one still finds an expectation value 
for the throat number of approximately $0.23$. In other words, generating 
the electroweak hierarchy is also very plausible since about 1 in 4
vacua have a sufficiently long throat. However, we can clearly not claim 
that throats of this length are unavoidable. 

We now turn to the case $K/3c=200$, where things look very different 
indeed. The longest expected throat produces a huge hierarchy $\sim 10^{80}$. 
Thus, we expect almost conformal field theories with very low IR cutoff (which 
are presumably only gravitationally coupled to standard model matter) to 
be abundant. Specifically, not having a throat with hierarchy $10^{29}$ or 
larger (corresponding to an IR scale of meV) has probability of about 5\%
(cf. Eq.~(\ref{p0})). In other words, hidden sectors with dynamical scales 
$\sim$meV or below are a {\it prediction} of this setting. Clearly, the above 
can have very important cosmological implications as far as dark matter or 
dark radiation are concerned. Just to give one more numerical implication 
of the formulae of the last section: The expected number of throats with a
hierarchy larger than $M_P/M_{EW}\sim 10^{14}$ is $\bar{n}\sim 6$. Thus, 
several electroweak-scale hidden sectors are a natural occurrence. The 
phenomenological and cosmological implications of this scenario clearly 
depend very strongly on whether `we' are in the throat or on the Calabi-Yau, 
where inflation took place and how strongly throat sectors are coupled 
to each other and to light fields localized in the UV. 
Away from the cosmological context there are two outstanding possibilities
for signatures of long throats with IR scale at or below the weak scale which
have been partially investigated: invisible Higgs decays to hidden sector 
particles\cite{Patt:2006fw}, and kinetic mixing of hypercharge with 
hidden-sector U(1)'s\cite{Kumar:2006gm}.  All we can say at 
present is that the various scenarios of this type studied in the 
literature appear to be everything else but exotic.

\section{Conclusions}
Based on a number of assumptions, we have quantified the expectation that 
throats are common in the type IIB landscape. The crucial starting point 
is the large number of 3-cycles which the compact space is expected to have. 
This can be quantified in two ways: conservatively, by taking the minimal
number which allows for the fine-tuning of $\Lambda$, or more optimistically, 
by taking a number which is typical for the more complicated Calabi-Yau 
manifolds. Given this large number of cycles (all of which generically 
carry a certain discrete flux number), one has to expect that by pure 
chance the flux on some of these cycles will be relatively small. Those
cycles are stabilized at small size, which generically leads to the 
development of a throat and a large hierarchy of scales. We have made this 
last argument more precise on the basis of the known behaviour of the 
density of flux vacua near conifold points. 

Our main technical results are simple formulae for the expectation value of
the number of throats with a certain hierarchy and for the probability of 
having no throat with a hierarchy larger than some given value. The
numerical predictions depend on the uncertain total number of 3-cycles 
mentioned above and on the details of the flux distribution away from the 
conifold points. Even with conservative assumptions about both of these 
unknown quantities, short throats (with hierarchies $\sim 10^3$) are 
generically expected while longer throats (with electroweak hierarchy) 
are at least not uncommon. Taking optimistic values for the unknown 
input data, we find that extreme hierarchies $\sim 10^{80}$ are expected and 
throats with electroweak hierarchy represent a firm statistical prediction. 

While these findings confirm the claim of our title that throats are 
ubiquitous in the type IIB landscape, the typical length of those throats
is quite uncertain at present. In this respect the main open questions are 
how complex a Calabi-Yau we should be looking for and a quantitative 
understanding of the `bulk' of the high-dimensional moduli space of such
manifolds. Furthermore, a better understanding of the role played by the 
large complex structure regions (which we have ignored in this analysis)
is highly desirable. 

Finally, even though many important questions remain unanswered, we consider 
one conclusion as relatively firm: Throats are common in the presently best
understood part of the string-theory landscape and should thus be taken 
very seriously both in string-theoretic and phenomenological model building. 
Given the very general setting we have been working with and the small
number of assumptions that we had to make, we are optimistic that 
throats will become one of the most firm and concrete predictions of the 
type IIB landscape. We would like to view this as strong support for 
phenomenological research in 5d Randall-Sundrum-like model, put within the 
more specific limits of their type IIB realization~\cite{Gherghetta:2006yq}. 
At the same time one should, however, keep in mind that, if the `favourable'
scenario of many very long throats is confirmed and cosmological problems 
with the various light fields are established, this whole line of thinking may 
turn into a serious argument against the type IIB landscape. 

\noindent
{\bf Acknowledgements}:\hspace*{.5cm}We would like to thank Savas Dimopoulos 
and Maximilian Kreuzer for helpful discussions.  We thank the Galileo Galilei
Institute for Theoretical Physics for hospitality and the INFN for partial
support during the completion of this work.


\end{document}